\documentstyle[epsf,letter,wrapft]{ptptex}

\begin{document}

\title{Hawking Radiation in the Laboratory}

\author{Masa-aki {\sc SAKAGAMI}\footnote
{E-mail: sakagami@phys.h.kyoto-u.ac.jp} and 
    Akira {\sc OHASHI}\footnote{E-mail: ohashi@cosmos.phys.ocha.ac.jp}}
\inst{$^*$Department of Fundamental Science, FIHS, 
Kyoto University, Kyoto 606-8501, Japan
\\
$^{**}$Department of Physics, Ochanomizu University, Tokyo 112-0012,
Japan
}


\recdate{
September 18, 2001}
\notypesetlogo

\abst{
We propose an experimental model using the Laval nozzle
of a sonic analogue of Hawking radiation.
We derive the power spectrum of the outgoing wave 
emitted from the vicinity of the sonic horizon 
instead of the created particle number.
Our treatment is based on classical theory, 
and it should make experiments easier to implement.
This experimental feasibility is a great advantage of 
our model.
}


\maketitle

Black hole evaporation, so-called Hawking radiation,
 is one of the most surprising predictions in theoretical physics.
\cite{hawking1,hawking2}
Despite the importance of Hawking radiation, 
its actual observation in the universe is considered to be very 
difficult. For this reason, it seems quite natural to seek 
experiments in the laboratory that can simulate this interesting 
phenomenon.   
In addition,
Unruh showed that there exists a sonic analogue of black holes.
We speculate that something like Hawking radiation 
may be observed in the evaporation of these analogous structures.
\cite{unruh1}\cite{unruh2}

Of course, black hole evaporation is a quantum phenomenon. 
Thus, in the case of the corresponding laboratory experiments,
we should maintain the quantum coherence of the entire system to 
observe processes of particle creation and changes of vacuum 
caused by the existence of the horizon.
With this in mind, several experiments employing  
superfluid  ${}^3$He have been proposed.
\cite{visser}\cite{volovik} 
However, the supersonic motion of this fluid along 
the walls of the apparatus would cause the 
collapse of superfluidity and quantum coherence,
because vortices may be created near the boundaries.
Additionally, even in the an experimental system\cite{volovik} 
designed to overcome this problem, no  
evidence of Hawking radiation has yet been reported.

In consideration of the points raised above,
we propose an experiment to observe 
a classical analogue of Hawking radiation,\cite{pad}
i.e. the power spectrum of an outgoing wave emitted from 
the vicinity of the horizon. 
Since we are only concerned with classical wave 
propagation in a fluid flow that has a sonic horizon, 
this experiment has nothing to do with quantum particle 
creation. Thus, we cannot detect the emission of particles 
corresponding to the evaporation of a black hole. 
However, Hawking radiation possesses another striking 
feature, a thermal spectrum. 
It can be shown that the power spectrum of 
an outgoing wave obeys the Planck distribution\cite{pad} 
and a role corresponding to that of the particle number 
in the usual scenario of black hole evaporation.
Because the behavior we describe does not require
 a quantum coherence of the system, the experiment 
we propose is much easier to carry out than 
the previous ones using ${}^3$He.

We first give an outline of the sonic 
analogue of the black hole in an ordinary fluid.
\cite{unruh1,unruh2}
We begin with a perfect fluid, which obeys 
\begin{eqnarray}
\rho\left(\displaystyle\frac{\partial \vec{v}}{\partial t}+(\vec{v}\cdot\nabla)\vec{v}\right)
    =-\nabla p(\rho),&\label{eq:euler}{~~}
\frac{\partial\rho}{\partial t}+\nabla(\rho\vec{v})=0,\label{eq:cont}&
\end{eqnarray}
where $\rho$, $p$ and $\vec{v}$ are the density, the pressure and
the velocity of the fluid.
Moreover, we assume the case of an adiabatic and rotation-free 
fluid, implying 
$
p=C\rho^{\gamma},~~\gamma=C_p/C_v,~~\nabla\times\vec{v}=0~({\rm i.e.}~~\vec{v}=\nabla\Phi),
$
where $C_p$ and $C_v$ are constant pressure and constant volume 
specific heat respectively, and
$\Phi$ is the velocity potential.
The Euler equation (\ref{eq:euler}) can be integrated, yielding
the Bernoulli's equation,
\begin{equation}
\frac{\partial \Phi}{\partial t} + 
\frac{1}{2}\vec{v}\cdot\vec{v} + h(\rho) = 0~~~.
\label{eq:Bernoulli}
\end{equation}
Here, $h(\rho)$
is the enthalpy and $c_s = \sqrt{(dp/d\rho)_{ad}} 
= \sqrt{\gamma~ p/\rho}$ is the local velocity of sound.
Then, we consider $\rho$ and $\Phi$ to be perturbated forms written
$\rho=\rho_0(1+\psi)$ and $\Phi=\Phi_0+\phi$
where $\rho_0$ and $\Phi_0$ are the density and velocity 
potential of the background flow.
After substituting these into (\ref{eq:cont}) and 
(\ref{eq:Bernoulli}),
we obtain the following equations for the perturbations:
\begin{eqnarray}
&\frac{\displaystyle \partial\phi}{\displaystyle \partial t}+
    \vec{v}_0\cdot\nabla\phi+c_s^2\psi=0,& \label{eq:per1} \\
&\frac{\displaystyle \partial\psi}{\displaystyle \partial t}+
    \vec{v}_0\cdot\nabla\psi
    +\triangle\phi+\nabla(\ln\rho_0)\cdot\nabla\phi=0~~~.
\label{eq:per2}&
\end{eqnarray}
The background flow obeys the same equations (\ref{eq:euler}), 
but with $\rho_0$ and $v_0$ replacing $\rho$ and $v$. 
Eliminating $\psi$ in
the Eqs. (\ref{eq:per1}) and (\ref{eq:per2}), we derive
the equation for the velocity potential $\phi$ as
\begin{eqnarray}\label{eq:wave-eq}\displaystyle
\left[\frac{1}{\rho_0}\left(\frac{\partial}{\partial t}+
\nabla\cdot\vec{v}_0\right)\frac{\rho_0}{c_s{}^2}
\left(\frac{\partial}{\partial t}+\vec{v}_0
\cdot\nabla\right)-\frac{1}{\rho_0}\nabla(\rho_0\nabla)\right]\phi=0.
\end{eqnarray}
This equation can be interpreted as describing
a scalar field $\phi$ on the background geometry
with the metric
\begin{equation}\label{eq:perturb}
g_{\mu \nu }  = \frac{{\alpha \rho _0 }}{{c_s }}\left( {\begin{array}{*{20}c}
   { - \left( {c_s{}^2  - \vec v_0 \cdot \vec v_0} \right)} & 
{ - v_0^i }  \\
   { - v_0^j } & {\delta _{ij} }  \\
\end{array}} \right),
\end{equation}
where $\alpha$ is a constant.
In order to understand the correspondence between
this geometry and a black hole,
let us consider one-dimensional flow
with  velocity $\vec{v_0}=(v_0(x),0,0)$.
In the $t$-$x$ plane, the metric can be
reduced to the  form
\begin{eqnarray}\displaystyle
ds^2=\frac{\alpha\rho_0}{c_s} 
\left[-\left(1-\frac{v_0(x)^2}{c_s(x)^2}\right)c_s(x)^2d\tau^2+
\frac{dx^2}{1-\frac{v_0(x)^2}{c_s(x)^2}}\right]~~,
\end{eqnarray}
where we have introduced the new time coordinate 
$\tau$ defined by
$d\tau = dt + \frac{v_0 dx}{c_s^2 - v_0^2}$.
This implies that the sonic point, at which $v_0(x)=c_s(x)$,
may play the role of the horizon.
By the same argument as in the case of the original Hawking 
radiation, it can be shown that 
this fluid model describes  Hawking radiation with 
a Planckian distribution,
and we can derive the ``surface gravity'' which determines
the temperature of the radiation as
\begin{equation}\displaystyle\label{eq:sg}
\kappa=  \left. \frac{d(c_s - v_0)}{dx}\right|_{\rm sonic~horizon}.
\end{equation}

\begin{wrapfigure}{r}{42mm}
    \epsfxsize=40mm
    \epsfysize=30mm
    \centerline{\epsfbox{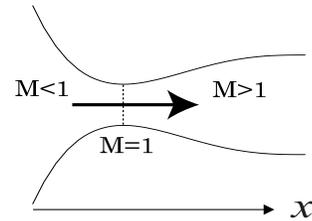}}
    \caption{Laval nozzle.}
    \label{fig:Laval}
\end{wrapfigure}

Here we propose a model using a Laval nozzle\cite{text}
that realizes the situation discussed above.
Consider an axi-symmetric tube with cross section $A(x)$, 
where $x$ is the coordinate along the tube.
When $A(x)$ changes only slightly along $x$, we can regard 
the flow to be uniform in cross section 
and  one dimensional.
For  stationary background flow along the Laval nozzle, 
from the Eqs. (\ref{eq:cont}) and (\ref{eq:Bernoulli}) 
we can derive a relation between 
the velocity of the flow and the area of the cross section
\begin{equation}\displaystyle
\left(M^2-1\right)\frac{dv_0}{v_0}=\frac{dA}{A},
\end{equation}
where $M=v_0/c_s$ is the Mach number.
This suggests that
subsonic flow will be accelerated when the cross section
becomes smaller,
and supersonic flow will be accelerated when 
the cross section increases.
In this paper, we consider a tube that has narrow throat
near the center of the tube and increases at the ends.
Such a tube is called a ``Laval nozzle''.
In a Laval nozzle,
$A(x)$ decreases along the direction of flow ($dA<0$)
in the region upstream from the throat and
increases ($dA>0$) in  the region downstream (Fig.\ref{fig:Laval}).
Therefore, if we can realize $M=1$ in the throat region,
we can realize  supersonic flows in the down stream region
and a sonic horizon emerges in the  
throat.
Suppose that there exist an asymptotic region in the upstream 
of the flow where the fluid is at rest, i.e. $v_0 = 0$ and 
the pressure and density are given by 
$p_0 = p_u,~ \rho_0 = \rho_u$. 
It is obvious that this region can be related to an asymptotically 
flat spacetime region in the corresponding black hole spacetime. 
Then  Bernoulli's equation (\ref{eq:Bernoulli}) becomes
\begin{equation}\label{eq:Bernoulli2}
\frac{1}{2}v_0^2 + \frac{\gamma}{\gamma-1}\frac{p_0}{\rho_0}=
\frac{\gamma}{\gamma-1}\frac{p_u}{\rho_u}
\end{equation}
which determines the velocity of the fluid as a function of the 
density $\rho_0/\rho_u$.

\begin{figure}[htb]
  \parbox{\halftext}{
    \epsfxsize=65mm
    \epsfysize=40mm
    \epsfbox{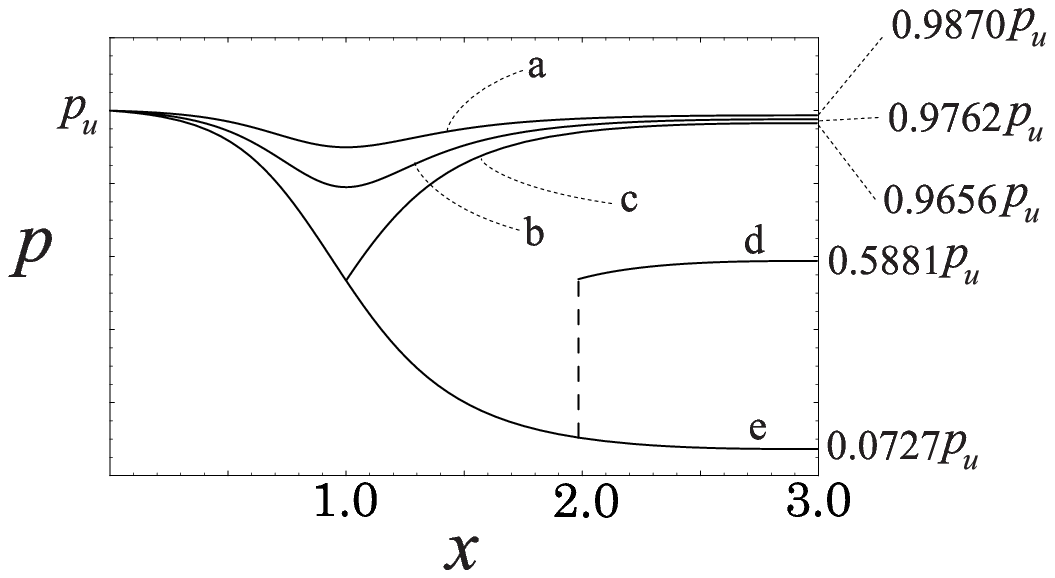}
  }
  \hspace{8mm}
  \parbox{\halftext}{
    \epsfxsize=65mm
    \epsfysize=40mm
    \epsfbox{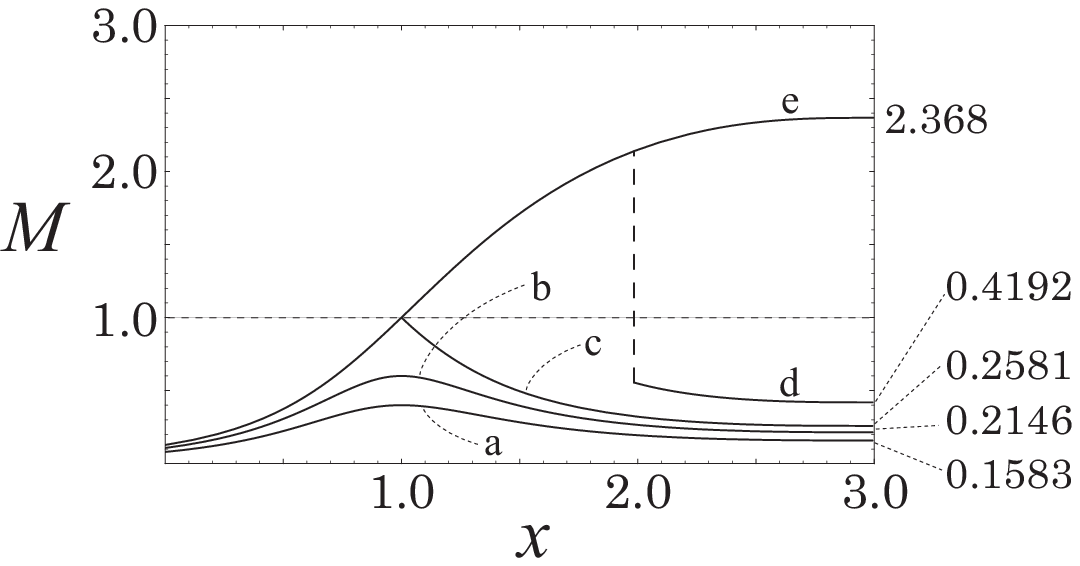}
  }
\caption{Profiles of the pressure $p$ and Mach number $M$ of 
the background flows along the Laval nozzle (Fig.\ref{fig:Laval}) 
in the case of $\gamma = 7/5$ . 
The inlet, throat and outlet of the nozzle are located 
at $x=0, 1$, and $3$, respectively. 
The vertical dashed-line indicates the shock surface. 
Above the critical value of the outlet pressure (case c), the flow 
along the nozzle is completely subsonic (cases a and b). Contrastingly,
the sonic horizon ($M=1$ at $x=1$) and regular 
supersonic flow are realized for sufficiently low 
pressure at the exit (case e). In the intermediate case (case d), 
a shock is formed in the upstream region, which does not affect   
the flow near the sonic horizon at $x=1$.  
}
\label{fig:profile}
\end{figure}

Figure \ref{fig:profile} displays profiles of the pressure and Mach 
number along the nozzle whose shape is depicted in Fig.\ref{fig:Laval}.
\cite{text} Under the condition that the inlet pressure at $x = 0$
is fixed,  the nature of the flow is determined by 
the value of the outlet pressure at $x=3$. 
We note that a sonic horizon at the throat can be formed 
without any fine-tuning, so that the horizon is 
stable with respect to changes in the parameter values of the nozzle. 
Furthermore, even in the unsteady situation, 
experimental and numerical studies of 
time-dependent flow within the nozzle\cite{Ott} show that the flow 
is settled to the above static one which is predicted by the Bernoulli 
equation (\ref{eq:Bernoulli2}).

%


Next, let us consider the fluctuations of the fluid.
Substituting $\phi(t,x)= e^{-i\omega t}$  ~~ $\times e^{i\int k(x)dx}$ into
Eq. (\ref{eq:wave-eq}) and using the WKB approximation,
we obtain 
\begin{equation}
( c_s{}^2 - v_0^2 )\, k^2 - 2\, \omega\, v_0\, k - \omega^2 = 0,
\label{eq:dispersion}
\end{equation}
whose solutions are given by
$
k = \omega/( c_s  - v_0) \equiv k_{\rm out}$  and 
$ 
k = - \omega/(c_s  + v_0) \equiv k_{\rm in}.
$
The outgoing solution $k_{\rm out}$ represents 
a wave that propagates upstream against the background flow.   
In the asymptotic region ($v_0 \sim 0$), this mode is a 
usual outgoing plane wave and should be regarded 
as an observable. However,
we note that this solution corresponds to a wave that 
marginally escapes from the sonic horizon $v_0 =  c_s$, 
so that the behavior near the horizon differs drastically 
from that in the asymptotic region.    
Contrastingly, $k_{\rm in}$ is an ingoing wave 
that propagates downstream and goes through
the horizon almost senselessly.
Hereafter, we are mainly concerned with the outgoing 
solution $k_{\rm out}$.

To proceed with the  investigation, 
we assume that the background flow of the fluid
satisfies $\rho_0/\rho_u=g(x/L)$ and $p_0/p_u=g^{\gamma}(x/L)$, where
$L$ is the characteristic length scale of the Laval nozzle.  Although 
the function $g$ is determined by the actual shape of 
the Laval nozzle, 
its precise form is not necessary in the subsequent 
analysis. We specify only the locations of the throat and the 
asymptotic region as follows: 
The throat is located 
at $x_{\rm th}$ (where ``th'' means ``throat''),
which satisfies $g(x_{\rm th}/L)=(2/(\gamma+1))^{1/(\gamma-1)}$,
and the asymptotic region in the upstream region is 
given by $x \rightarrow +\infty, v_0 \rightarrow 0 $.
Under these conditions and using Eq. (\ref{eq:Bernoulli2}), 
we obtain
\begin{equation}
\frac{c_s-v_0}{c_{su}}=g^{\frac{\gamma-1}{2}}\left(\frac{x}{L}\right)-
\sqrt{\displaystyle\frac{2}{\gamma-1}
\left(1-g^{\gamma-1}\left(\frac{x}{L}\right)\right)}.
\end{equation}
When we introduce a coordinate $z$ near the sonic 
point $x_{\rm th}$ as $x=x_{\rm th}+z$ ,
we can calculate the surface gravity using 
Eq.(\ref{eq:sg}), 
\begin{equation}\label{eq:surfacegravity}
\kappa=\frac{g'_{\rm th}}{L}
\left(\frac{\gamma+1}{2}\right)^{(\gamma-3)/2(\gamma-1)},
~~g'_{\rm th} = \left.\frac{dg(y)}{dy}\right|_{y=(x_{\rm th}/L)}.
\end{equation}
Thus, we obtain the following expression for the outgoing mode of 
the perturbations 
near the sonic horizon, $x \sim x_{\rm th}$ 
\begin{equation}
\phi^{{\rm out}}_{\omega}=
\exp\left(i\frac{\omega}{c_s\kappa}\ln z\right)=
z^{i\omega/c_s\kappa}.
\label{eq:log-wave}
\end{equation}
We note that this wave behaves as a usual out-going plane wave 
\begin{equation}
\phi^{{\rm out}}_{\omega}=\exp\left(i\frac{\omega}{c_{su}}x\right)
\label{eq:usual-out}
\end{equation}
in the asymptotic region, i.e. $x\rightarrow +\infty$.

\begin{figure}[b]
    \epsfxsize=90mm
    \epsfysize=41mm
    \centerline{\epsfbox{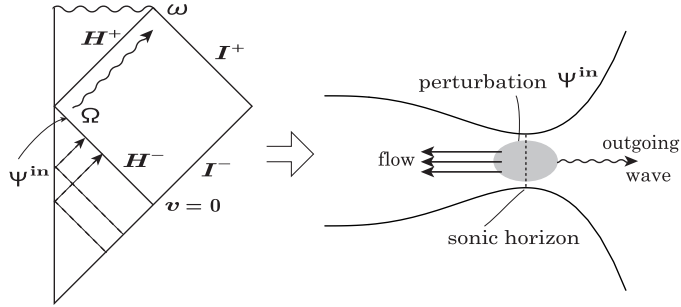}}
    \caption{Comparison between the black hole spacetime 
and the flow of the Laval nozzle.}
    \label{fig:compare}
\end{figure}

In Fig. \ref{fig:compare}, we give a  comparison between 
the usual black hole spacetime and the fluid flow in 
the Laval nozzle. In the scenario of Hawking radiation,
\cite{hawking1,hawking2}\cite{review} a normal mode  
corresponding to the vacuum, 
i.e. a positive frequency part, is prepared in the past 
null infinity $I^-$. Here, we consider the evolution of 
this mode toward the past horizon $H^-$. 
The existence of the black hole region and 
the future horizon $H^+$ do not affect this 
evolution so that a quantum state on $H^-$ 
is nothing more than the vacuum state in $I^-$.  
An observer prepares 
the vacuum state in the future null infinity $I^+$,
which is determined by a positive frequency normal mode in $I^+$, 
and he defines a particle in the vacuum.
As in the previous case, we investigate 
the evolution of the mode in $I^+$ {\it backward in time} 
toward the past horizon $H^-$. 
In this case, however, the future horizon $H^+$ 
significantly affects the evolution of the mode. 
Thus, a comparison between the normal modes in $I^-$ and $I^+$ 
in the same place $H^-$ 
shows that the future vacuum in $I^+$ differs from 
the past one in $I^-$, and the thermal emission of 
particle can be observed in the future null infinity $I^-$. 

We note that the propagation of the perturbation 
in the background flow in the Laval nozzle possesses 
an essential feature of Hawking radiation in black hole 
spacetime.  
The outgoing plane wave (\ref{eq:usual-out}) 
in the asymptotic region, which corresponds to 
the normal mode in the future null infinity $I^+$, 
is affected significantly by the sonic horizon,
and therefore, its behavior near 
the sonic horizon is drastically changed, as described by  
(\ref{eq:log-wave}). It is well known that 
the logarithmic behavior of the mode (\ref{eq:log-wave})
results in the thermal property of the spectrum of 
the created particles and its temperature is 
related to the value of the surface gravity $\kappa$. 
This means that if we could realize the ideal system 
in which the quantum coherence is being maintained perfectly,
we would be able to simulate  Hawking radiation in 
the fluid in a Laval nozzle.
However, it seems that this would be a very difficult experiment
to carry out with a sonic horizon, because interaction of 
the fluid with the wall of the apparatus would cause the vortex formation
and destroy the quantum coherence of the  system.

Because of the difficulty described above, 
we treat our problem classically and look 
for the classical analogue of  Hawking radiation.\cite{pad} 
We prepare an outgoing wave at the past (sonic) horizon $H^-$ as
\begin{equation}
\Psi^{\rm in}_{\Omega}=\exp(i\Omega z) 
\end{equation}
which mimics the vacuum state function in the past null infinity $I^-$.
Suppose that we observe this wave in the asymptotic region $I^+$ 
in terms of a {\it power spectrum} instead of  
the particle number. 
For this purpose, we evaluate  the Fourier components  
\begin{eqnarray}
f(\omega)=\int^{\infty}_{0}dz{~}\Psi^{\rm in}_{\Omega}
\phi^{\rm out *}_{\omega}
=- i\Omega^{-(1-\frac{i\omega}{c_s\kappa})}~
\Gamma(1-\frac{i\omega}{c_s\kappa})~e^{-\frac{\pi\omega}{2c_s\kappa}}
\end{eqnarray}
of the prepared wave $\Psi^{\rm in}_{\omega}$ 
with respect to the observed wave in 
the asymptotic region $\phi^{\rm out}_\omega$. 
Finally, we obtain the power-spectrum that would be observed
upstream:
\begin{equation}
\left|f(\omega)\right|^2 \propto \frac{2\pi\omega}{c_s\kappa}
\frac{1}{e^{2\pi\omega/c_s\kappa}-1}.
\label{eq:Planck}
\end{equation}
This is the Planckian distribution, 
which can be regarded as the classical analogue of 
Hawking radiation. From Eqs.(\ref{eq:surfacegravity}) 
and (\ref{eq:Planck}), the spectrum 
is characterized by a wavelength $\lambda \sim 1/\kappa$. 

Of course, our proposal does not represent 
a full quantum simulation of Hawking radiation. 
However, the nature of the wave propagation near the horizon  
in the presently considered system is similar to that in 
the case of Hawking radiation. 
In particular, we should be able to 
observe a Planckian distribution,
which is one of the important features of Hawking radiation.
One of the main advantages of the present system is that 
actually carrying out an experiment
and observing the power spectrum would be much easier 
than the previous proposal.\cite{volovik}
In the present case, the sonic horizon is believed to be stable.
In addition to  experimental and numerical results\cite{Ott}
that indicates this stability, 
our result (\ref{eq:dispersion}) itself, which  indicates 
the non-existence of growing modes in the perturbation equation 
(\ref{eq:wave-eq}), can be recognized as  proof of the stability. 
Taking into account these advantages, 
we  regard our proposal as a first step toward 
simulation of  Hawking radiation 
in the laboratory. 


%
%

\end{document}